\documentstyle[preprint,aps]{revtex}

\newcommand{\ba}{\begin{eqnarray}}
\newcommand{\ea}{\end{eqnarray}}
\begin{document}
\pagestyle{plain}
\title{Partial Dynamical Symmetry in Deformed Nuclei}
\author{
Amiram Leviatan\\
Racah Institute of Physics, The Hebrew University,\\
Jerusalem 91904, Israel}
\maketitle
\begin{abstract}
We discuss the notion of partial dynamical symmetry in relation to nuclear
spectroscopy. Explicit forms of Hamiltonians with partial $SU(3)$ symmetry
are presented in the framework of the interacting boson model of nuclei.
An analysis of the resulting spectrum and electromagnetic transitions
demonstrates the relevance of such partial symmetry to the spectroscopy of
axially deformed nuclei. 
\end{abstract}
\begin{center}
\end{center}

\newpage

Recent years, in particular since the introduction of the interacting boson
model of nuclei (IBM) \cite{IBM}, have witnessed substantial progress in
developing algebraic symmetry-based models, which are now part of the
standard lexicon of nuclear structure \cite{ALG}.
A characteristic and attractive feature in these models is the occurrence
of dynamical symmetries. This corresponds to a situation in which the
Hamiltonian is written in terms of Casimir operators of a chain of nested 
groups. A dynamical symmetry provides considerable insight since
it allows all properties of the system 
to be calculated in closed form.
The labels of irreducible representations (irreps) of 
the groups in the chain serve as quantum numbers to classify members of
a complete basis in which the Hamiltonian is diagonal. 
The group-theoretical classification scheme inherent
to the dynamical symmetry basis facilitates the numerical treatment and
interpretation of the general Hamiltonian.
 
The merits of having a (dynamical) symmetry are self-evident.
However, in detailed applications of group theoretical schemes to the
spectroscopy of nuclei, one often finds that the assumed symmetry is
not obeyed uniformly, i.e. some levels fulfill the symmetry while other
levels do not. Exact symmetries impose severe constraints on the
corresponding spectrum (e.g. particular band degeneracies) which are rarely
observed in real nuclei. These observations motivate
one to consider a particular symmetry- breaking that would result in 
mixing of irreps in some part of the spectrum while retaining a
good symmetry to specific eigenstates. We refer to such a situation as
partial (dynamical) symmetry. 
Within such symmetry construction only a subset of eigenstates
are pure and preserve the desired features of a dynamical symmetry.
IBM Hamiltonians with F-spin partial symmetry were shown in \cite{LGK}.
The mathematical aspects and algorithm for 
partial dynamical symmetries (pds) were presented in \cite{PDS}. 
The purpose of the present work is to show that
pds are not just a formal mathematical notion but rather
are actually realized in nuclei and thus
may serve as a useful tool in realistic applications of algebraic methods 
to nuclear spectroscopy. In this letter we consider $^{168}$Er as a typical
example of an axially deformed prolate nucleus in the rare earth region,
and show the relevance of $SU(3)$ pds to its description.

The starting point for the IBM description of
axially deformed nuclei is the $SU(3)$ dynamical symmetry, 
corresponding to the chain $U(6) \supset SU(3) \supset O(3)$.
The basis states are labeled by $\vert [N](\lambda,\mu)KLM\rangle$, 
where $N$ is the total number of monopole $(s^{\dagger})$ and
quadrupole $(d^{\dagger})$ bosons, $L$ the angular momentum, 
$(\lambda,\mu)$ denote the $SU(3)$ irreps and $K$ is an additional label
needed for complete classification and corresponds geometrically to the
projection of the angular momentum on the symmetry axis. 
The Hamiltonian in this case
involves a linear combination of the Casimir operators of $SU(3)$ and
$O(3)$. The corresponding eigenstates are arranged in $SU(3)$ multiplets.
The lowest $SU(3)$ irrep is $(2N,0)$ which describes the ground band $g(K=0)$
of an axially deformed nucleus.
The first excited $SU(3)$ irrep
$(2N-4,2)$ contains both the $\beta(K=0)$ and $\gamma(K=2)$ bands. 
Consequently, states in these bands with the same angular momentum
are degenerate.
This $\beta$-$\gamma$ degeneracy is a characteristic feature of the $SU(3)$
limit of the IBM which, however, is not commonly observed \cite{CW}.
In most deformed nuclei the $\beta$ band lies above the $\gamma$ band
as is evident from the experimental spectrum of $^{168}$Er shown in Fig. 1. 
In the IBM framework, with at most two-body interactions, one
is therefore compelled to break $SU(3)$ 
in order to conform with the experimental data. 
To do so, the usual approach has been to include in the Hamiltonian
terms from other chains
so as to lift the undesired $\beta$-$\gamma$ degeneracy.
Such an approach was taken in ref. \cite{WCD} where
an $O(6)$ term was added to the $SU(3)$ Hamiltonian yielding
a satisfactory description of
the spectroscopic data of $^{168}$Er below 2 MeV, as shown in Fig. 1. 
However, in this procedure,
the $SU(3)$ symmetry is completely broken,
all eigenstates are mixed and no analytic solutions are retained. 
Similar statements apply to the description in the consistent Q formalism 
\cite{CQF}. In contrast, partial $SU(3)$ symmetry, to be discussed below, 
corresponds to breaking $SU(3)$, but in a very particular way
so that {\bf part} of the states (but not all) will still be solvable with
good symmetry. As such, the virtues of a dynamical symmetry 
(e.g. solvability) are fulfilled but by only a subset of states. 

To consider partial $SU(3)$ symmetry in the IBM framework we examine
the following rotational-invariant Hamiltonian
\ba
H(h_0,h_2) &=& h_{2}\left [-\hat C_{SU(3)} + 2\hat N (2\hat N+3)\right ]
\nonumber\\
&& +\, (h_2-h_0)\left [ -4\hat N^2 - 6\hat N + \hat n_d - \hat n_d^2 
+ 4\hat N\hat n_d + 2\hat C_{O(6)} -\hat C_{O(5)}\right ]\qquad
\ea
where $h_{0}, h_{2}$ are arbitrary constants and we use the definition
of Casimir operators as in Table I of the Appendix in ref. \cite{LEV}.
Clearly, for $h_{0}\neq h_{2}$ the above Hamiltonian contains a mixture
of Casimir operators of all IBM chains, hence it breaks the $SU(3)$
symmetry. However, it respects $SU(3)$ as a partial 
symmetry. To confirm this non-trivial statement, it is simpler to
consider the normal order form \cite{LEV,LK}
\ba
H(h_0,h_2) \;=\; h_{0}P^{\dagger}_{0}P_{0} + h_{2}P^{\dagger}_{2}
\cdot\tilde P_{2}
\quad ~,
\ea
where $\tilde P_{2,\mu} = (-)^{\mu}P_{2,-\mu}$. 
The Hamiltonian is seen to
be constructed from boson pair operators
with angular momentum $L=0$ and $2$, which are defined as
\ba
P^{\dagger}_{0}\; &=& \; d^{\dagger}\cdot d^{\dagger} - 2(s^{\dagger})^2
\;\;\; , \;\;\;
P^{\dagger}_{2,\mu}\; = \; 2\,s^{\dagger}d^{\dagger}_{\mu} 
+ \sqrt{7}(d^{\dagger}d^{\dagger})^{(2)}_{\mu} \quad ~.
\ea                              
These boson pair operators satisfy the following properties
\begin{eqnarray}
&&P_{L,\mu}\vert c;N\rangle \;=\; 0
\;\;\; , \;\;\;
\left [P_{L,\mu}\, ,\, P^{\dagger}_{2,2}\right ]\vert c;N\rangle\;= \;
\delta_{L,2}\delta_{\mu,2}\,6(2N+3)\vert c;N\rangle \;\;\; , \qquad
\nonumber\\
&&\left [\left [P_{L,\mu}\, , \,P^{\dagger}_{2,2}\right ]\, , \,
P^{\dagger}_{2,2}\right ]\; = \;
\delta_{L,2}\delta_{\mu,2}\,24P^{\dagger}_{2,2}\;\;\; ,
\qquad\quad L=0,2
\quad ~.
\end{eqnarray}
The state $\vert c;N\rangle \propto
[(s^{\dagger}+\sqrt{2}d^{\dagger}_{0})]^{N}\vert 0\rangle$
in Eq. (4) 
is a condensate of N bosons 
which serves as an intrinsic state \cite{GK} for the
$SU(3)$ ground band. 
For arbitrary $h_{0}, h_{2}$ coefficients
the Hamiltonian $H(h_0,h_2)$ 
is not an $SU(3)$ scalar. Nevertheless, it has a subset
of eigenstates with good $SU(3)$ character. This follows from relations (4)
which imply that the sequence of states
\ba
\vert k\rangle \; \propto \; \left (P^{\dagger}_{2,2}\right )^{k}
\vert c; N-2k\rangle
\quad ~,
\ea
are eigenstates of $H(h_0,h_2)$ 
with eigenvalues $E_{k}\;=\; 6h_{2}\left (2N+1 -2k \right )k $.
These energies are the $SU(3)$ eigenvalues of $H(h_0=h_2)$, 
and identify the states $\vert k\rangle$ to be in the
$SU(3)$ irreps
$(2N-4k,2k)$ with $2k\le N$. It can be further shown that they are lowest
weight states in these representations.
The states $\vert k\rangle$ are deformed and serve as intrinsic states 
representing $\gamma^{k}$ bands with angular momentum projection ($K=2k$) 
along the symmetry axis \cite{CA}. In particular, $\vert k=0\rangle$
represents the ground-state band ($K=0$) and $\vert k=1\rangle$
is the $\gamma$-band ($K=2$). 
The intrinsic states break the $O(3)$ symmetry but since the Hamiltonian
in Eq. (2) is an $O(3)$ scalar, the projected states 
are also eigenstates of $H(h_0,h_2)$ with energy $E_k$ and with good 
$SU(3)$ symmetry. For the ground band $(k=0)$ the projected states span the
entire $SU(3)$ irrep $(2N,0)$. For excited bands $(k\neq 0)$, the projected 
states span only part of the corresponding
$SU(3)$ irreps. There are other states originally in these irreps 
(as well as in other irreps) which do
not preserve the $SU(3)$ symmetry and therefore get mixed.
In particular, the ground $(g)$ and $\gamma$ bands retain their $SU(3)$
character $(2N,0)$ and $(2N-4,2)$ respectively, but the $\beta$ band is mixed.
This situation corresponds precisely to that of partial $SU(3)$ symmetry.
An Hamiltonian $H(h_0,h_2)$ which is not an $SU(3)$ scalar has a subset of 
{\bf solvable} eigenstates which continue to have  good  $SU(3)$ symmetry. 
All of the above discussion is applicable also to the case when
we add to the Hamiltonian (2) the Casimir operator of $O(3)$ 
($\hat C_{O(3)}$), and by doing so convert the partial $SU(3)$ symmetry
into partial dynamical $SU(3)$ symmetry.
The additional rotational term contributes just an $L(L+1)$ splitting 
but does not affect the wave functions.

The experimental spectra \cite{WCD} of the ground ($g$), $\beta$, and
$\gamma$ 
bands in $^{168}$Er is shown in Fig. 1. We now attempt a description in 
terms of an IBM Hamiltonian with partial dynamical $SU(3)$ symmetry
\ba
H \;=\; H(h_0,h_2) + \lambda\, \hat C_{O(3)} \quad ~. 
\ea
According to the previous discussion, the spectrum
of the ground and $\gamma$ bands is given by
\ba
E_{g}(L) &=& \lambda L(L+1) \;\;\; , \;\;\;
E_{\gamma}(L) = 6h_{2}(2N-1) + \lambda L(L+1) \quad ~.
\ea
The Hamiltonian in Eq. (6) is specified by three parameters (N=16 for
$^{168}$Er according to the usual boson counting).
We extract the values of $\lambda$ and $h_{2}$ from the 
experimental energy differences
$[E(2^{+}_{g})-E(0^{+}_{g})]$ and $[E(2^{+}_{\gamma})-E(2^{+}_{g})]$ 
respectively.
For an exact $SU(3)$ dynamical symmetry, $h_{0}=h_{2}$, implying 
$E_{\beta}(L)=E_{\gamma}(L)$ for even values of $L\geq 2$. The corresponding
spectrum (shown in Fig. 1) deviates considerably from the experimental
data since empirically the $\beta$ and $\gamma$ bands are not degenerate. On
the other hand, when the dynamical $SU(3)$ symmetry is only partial, one can
vary $h_{0}$ so as to reproduce the $\beta$ bandhead energy $E_{\beta}(L=0)$.
Having determined the parameters $\lambda, h_0, h_2$ from three
experimental energies, the prediction
for other rotational members of the ground $\beta$ and $\gamma$ bands
is shown in Fig. 1. No further attempt to improve the agreement between theory
and experiment was made since the philosophy of this calculation was to
investigate the validity of the $SU(3)$ pds. 
Clearly, the $SU(3)$ pds spectrum is an
improvement over the schematic, exact $SU(3)$ dynamical symmetry 
description, since the $\beta$-$\gamma$ degeneracy is lifted.
The good $SU(3)$ character, $(32,0)$ for
the ground band and $(28,2)$ for $\gamma$ band, is retained in the pds 
calculation, while the $\beta$ band contains $10\%$ $(26,0)$ and 
$3\%$ $(24,4)$ admixtures into the dominant $(28,2)$ irrep.
The quality of the calculated pds spectrum is similar to that obtained
in the broken $SU(3)$ calculation \cite{WCD} also shown in Fig.~1.

Electromagnetic transitions are a more sensitive probe to the structure
of states, hence are an important indicator for verifying
the relevance of partial $SU(3)$ symmetry. To calculate such observables
we need to specify the wave-functions of the initial and final states
as well as the operator that induces the transition.
For the Hamiltonian in Eq. (6), with partial dynamical $SU(3)$ symmetry,
the solvable states are those projected from the intrinsic states 
$\vert k\rangle = \vert (\gamma)^{k} (2N-4k,2k)K=2k\rangle$ of Eq. (5), 
and are simply selected members
of the Elliott basis $\phi_{E}((\lambda,\mu)KLM)$ \cite{ELL}. 
In particular, the states belonging to the ground and $\gamma$ bands are the
Elliott states $\phi_{E}((2N,0)K=0,LM)$ and $\phi_{E}((2N-4,2)K=2,LM)$
respectively. Their wave functions can be expressed in terms of 
the Vergados basis $\Psi_{V}((\lambda,\mu)\chi LM)$ \cite{VER}, which
is the usual (but not unique) choice for orthonormal $SU(3)$ basis.
The most general IBM one-body E2 operator may be written as 
\ba
T(E2) \; = \; \alpha\, Q^{(2)} + \theta\, (\,d^{\dagger}s 
+ s^{\dagger}\tilde d \,)
\ea
where $Q^{(2)}$ is the quadrupole $SU(3)$ generator.
The matrix elements of such $E2$ operator in the Vergados basis 
are known 
\cite{SU3,ISA}. 
It is therefore possible to obtain
{\bf analytic} expressions for the E2 rates
between the subset of solvable states. 
For the ground band and for members of the $\gamma$ band with $L$ odd, the
Vergados and Elliott bases are identical. Accordingly, 
the corresponding B(E2) values in the two bases are the same.
The Elliott states in the
$\gamma (K=2)$ band with even
values of $L$ are mixtures of Vergados states in the $\beta (\chi=0)$
and $\gamma(\chi=2)$ bands. The corresponding B(E2) value is 
\ba
&B_{E}(E2;\gamma K=2,L\rightarrow g K=0,L') =\qquad\qquad\qquad
\qquad\qquad\qquad
\qquad\qquad
\nonumber\\
&\qquad\qquad
\left [\,{\sqrt{B_{V}(E2;\gamma\chi=2,L\rightarrow 
g\chi=0,L')}
\;\pm\;x^{(L)}_{20}
\sqrt{B_{V}(E2;\beta\chi=0,L\rightarrow g\chi=0,L')}\over
x_{22}^{(L)}}\,\right ]^{2}
\ea
where the $+$ $(-)$ sign applies to a transition with $L'=L$ $(L'=L\pm 2)$.
In Eq. (9) the notation
$B_{V}(E2)$ and $B_{E}(E2)$ stands for $B(E2)$ values
calculated in the Vergados and Elliott bases respectively. 
The $x^{(L)}_{20},\, x^{(L)}_{22}$ are coefficients which appear in the
transformation between the two bases \cite{VER}.
Analytic expressions of $B_{V}(E2)$ values for $g\rightarrow g$ and
$\gamma\rightarrow g$ transitions have been derived \cite{SU3,ISA}. 

To compare with experimental data on $B(E2)$ ratios,
we adapt the procedure of ref. \cite{WCD} and extract
the parameters $\alpha$ and $\theta$ of the $E2$ 
operator in Eq. (8) from the experimental values of
$B(E2;0^{+}_{g}\rightarrow 2^{+}_{g})$ and 
$B(E2;0^{+}_{g}\rightarrow 2^{+}_{\gamma})$.
The corresponding ratio for $^{168}$Er is $\theta/\alpha=4.261$.
As shown in Table I, the resulting $SU(3)$ pds 
E2 rates for transitions originating within the $\gamma$ band 
are found to be in excellent agreement with
experiment and are similar to the calculation by Casten Warner and 
Davidson \cite{WCD} (where the $SU(3)$ symmetry is broken for all states).
In particular, the $SU(3)$ pds calculation reproduces correctly the ratio of
$(\gamma\rightarrow\gamma)/(\gamma\rightarrow g)$ strengths.
The only significant discrepancy is that for the $8^{+}_{\gamma}\rightarrow
7^{+}_{\gamma}$ transition which is very weak experimentally, with an
intensity error of $50\%$ and an unknown $M1$ component \cite{WCD}.
For transitions from the $\beta$ band the
overall agreement is good (better for $\beta\rightarrow\gamma$ transitions) 
although not as precise as for the
$\gamma$ band. The calculation exhibits the observed dominance of 
$\beta\rightarrow\gamma$ over $\beta\rightarrow g$ transitions.
As an example, 
for $2^{+}_{\beta}\rightarrow J_{f}$ transitions with
$J_{f}=(0^{+}_{g},4^{+}_{g},2^{+}_{\gamma},3^{+}_{\gamma},0^{+}_{\beta})$
the calculated and experimental $B(E2)$ ratios are 
$(0.42:1.44:2.59:4.77:100.0)$ and $(0.23:1.4:4.0:4.9:100.0)$ respectively.
A comparison with the prediction of an exact $SU(3)$ symmetry for these 
ratios: $(0.47,1.62,0.93,1.66,100.0)$ highlights the importance of 
$SU(3)$ mixing in the $\beta$ band. If we recall that only the ground
band has $SU(3)$ components $(\lambda,\mu)=(2N,0)$ 
and that $Q^{(2)}$ in Eq. (8) is a generator of $SU(3)$ (hence cannot
connect different $(\lambda,\mu)$ irreps), it follows that 
$\beta,\gamma\rightarrow g$ $B(E2)$ ratios are independent of both 
$\alpha$ and $\theta$. Furthermore, since the
ground and $\gamma$ bands have pure $SU(3)$ character, $(2N,0)$ and
$(2N-4,2)$ respectively, the corresponding wave-functions do not depend
on parameters of the Hamiltonian and hence are determined solely by
symmetry. Consequently, the 
B(E2) ratios for $\gamma\rightarrow g$ transitions quoted in Table I are 
parameter-free predictions of $SU(3)$ pds. 
The agreement between these predictions and the data
confirms the relevance of partial dynamical $SU(3)$ symmetry to the 
spectroscopy of $^{168}$Er.

To summarize, we have analyzed IBM Hamiltonians with $SU(3)$ pds. 
Such Hamiltonians are not
invariant under $SU(3)$ but have a subset of eigenstates with good $SU(3)$
symmetry. The special states are solvable and span part of particular 
$SU(3)$ irreps. Their wave-functions, eigenvalues and $E2$ rates are known
analytically. An application of the scheme to $^{168}$Er has demonstrated
that the empirical spectrum and $E2$ rates conform with the predictions
of partial $SU(3)$ symmetry. These observations point at the relevance of
partial $SU(3)$ symmetry to the spectroscopy of axially deformed nuclei,
at least as a starting point for further refinements. 

The notion of partial dynamical symmetry is not confined to $SU(3)$.
A general algorithm is available for constructing Hamiltonians 
with pds for any semi-simple group \cite{PDS}.
The occurrence of partial (but exact) symmetries imply that part of the
eigenvalues and wave functions can be found analytically but not the entire
spectrum. As such, pds can overcome the 
schematic features of exact dynamical symmetries (e.g.
undesired degeneracies) and simultaneously 
retain their virtues (i.e. solvability) for some states. 
We also wish to point out that 
Hamiltonians with partial symmetries are not completely integrable and may
exhibit chaotic behavior. This makes them a useful tool to study mixed
systems with coexisting regularity and chaos \cite{WAL}. 
It will be of great interest to explore the ramifications of partial
symmetries both for discrete spectroscopy and statistical aspects of nuclei.

This research was supported by the Israel Science Foundation administered
by the Israel Academy of Sciences and Humanities.

\newpage
 \begin{table}
 \caption[]{
$B(E2)$ branching ratios from states in the $\gamma$ band in
$^{168}$Er. The experimental ratios (EXP) and the broken $SU(3)$ calculation
of Warner Casten and Davidson (WCD) are taken from ref. \cite{WCD}. 
(PDS) are the partial dynamical $SU(3)$ symmetry calculation reported
in the present work.
\normalsize}
\vskip 20pt
 \begin{tabular}{lcccc|rlcccc}
                 &                  &          &          & & & & & & &  \\
 $J^{\pi}_{i}$ & $J^{\pi}_{f}$ &  EXP &  PDS &  WCD &    &
 $J^{\pi}_{i}$ & $J^{\pi}_{f}$ &  EXP &  PDS &  WCD \\
 \hline
                 &                  &          &          & & & & & & &  \\
$2^{+}_{\gamma}$ & $0^{+}_{g}$      & $54.0$   &  $64.27$ &  $66.0$  &    &
$6^{+}_{\gamma}$ & $4^{+}_{g}$      &   $0.44$ &   $0.89$ &   $0.97$ \\
                 & $2^{+}_{g}$      & $100.0$  & $100.0$  & $100.0$  &    &
                 & $6^{+}_{g}$      &   $3.8$  &   $4.38$ &   $4.3$  \\
                 & $4^{+}_{g}$      &   $6.8$  &   $6.26$ &   $6.0$  &    &
                 & $8^{+}_{g}$      &   $1.4$  &   $0.79$ &   $0.73$ \\
$3^{+}_{\gamma}$ & $2^{+}_{g}$      &   $2.6$  &   $2.70$ &   $2.7$  &    &
                 & $4^{+}_{\gamma}$ & $100.0$  & $100.0$  & $100.0$  \\
                 & $4^{+}_{g}$      &   $1.7$  &   $1.33$ &   $1.3$  &    &
                 & $5^{+}_{\gamma}$ &  $69.0$  &  $58.61$ &  $59.0$  \\
                 & $2^{+}_{\gamma}$ & $100.0$  & $100.0$  & $100.0$  &    &
$7^{+}_{\gamma}$ & $6^{+}_{g}$      &   $0.74$ &   $2.62$ &   $2.7$  \\
$4^{+}_{\gamma}$ & $2^{+}_{g}$      &   $1.6$  &   $2.39$ &   $2.5$  &    &
                 & $5^{+}_{\gamma}$ & $100.0$  & $100.0$  & $100.0$  \\
                 & $4^{+}_{g}$      &   $8.1$  &   $8.52$ &   $8.3$  &    &
                 & $6^{+}_{\gamma}$ &  $59.0$  &  $39.22$ &  $39.0$  \\
                 & $6^{+}_{g}$      &   $1.1$  &   $1.07$ &   $1.0$  &    &
$8^{+}_{\gamma}$ & $6^{+}_{g}$      &   $1.8$  &   $0.59$ &   $0.67$ \\
                 & $2^{+}_{\gamma}$ & $100.0$  & $100.0$  & $100.0$  &    &
                 & $8^{+}_{g}$      &   $5.1$  &   $3.57$ &   $3.5$  \\
$5^{+}_{\gamma}$ & $4^{+}_{g}$      &   $2.91$ &   $4.15$ &   $4.3$  &    &
                 & $6^{+}_{\gamma}$ & $100.0$  & $100.0$  & $100.0$  \\
                 & $6^{+}_{g}$      &   $3.6$  &   $3.31$ &   $3.1$  &    &
                 & $7^{+}_{\gamma}$ & $135.0$  &  $28.64$ &  $29.0$  \\
                 & $3^{+}_{\gamma}$ & $100.0$  & $100.0$  & $100.0$  &    &
                 &                  &          &          &          \\
                 & $4^{+}_{\gamma}$ & $122.0$  &  $98.22$ &  $98.5$  \\
                 &                  &          &          & & & & & & &  \\
 \end{tabular}
 \end{table}
\vspace{14.5pt}

\clearpage
\noindent
{\bf Figure 1:} Spectra of $^{168}$Er. Experimental energies
(EXP) are compared with an IBM calculation in an exact $SU(3)$ dynamical
symmetry ($SU(3)$), in a broken $SU(3)$ symmetry \cite{WCD} and in 
a partial dynamical $SU(3)$ symmetry (PDS). The latter employs the 
Hamiltonian of Eq. (6) with $h_0=0.008,\, h_2=0.004,\,
\lambda=0.013$ MeV. 


\newpage
\begin{thebibliography}{99}

\bibitem{IBM}
F.~Iachello and A.~Arima, {\it The Interacting Boson Model},
Cambridge Univ. Press, Cambridge, (1987).

\bibitem{ALG}
R.F. Casten in, {\it Algebraic Approaches to Nuclear Structure} 
(R.F. Casten Ed.) Harwood Academic Publishers (1993).

\bibitem{LGK}
A. Leviatan, J.N. Ginocchio and M.W. Kirson,
Phys. Rev. Lett. {\bf 65} (1990) 2853.

\bibitem{PDS}
Y. Alhassid and A. Leviatan, J. Phys. {\bf A25} (1992) L1265.\\
A. Leviatan, in {\it Symmetries in Science VII} (B. Gruber and T. Otsuka 
Eds.), Plenum Press, NY, (1992), p. 383.

\bibitem{CW}
R.F. Casten and D.D. Warner, Rev. Mod. Phys. {\bf 60} (1988) 389. 

\bibitem{WCD}
D.D. Warner, R.F. Casten and W.F. Davidson, Phys. Rev.
{\bf C24} (1981) 1713.

\bibitem{CQF}
D.D. Warner and R.F. Casten, Phys. Rev.
{\bf C28} (1983) 1798.

\bibitem{LEV}
A. Leviatan, Ann. Phys. {\bf 179} (1987) 201.

\bibitem{LK}
A. Leviatan, Z. Phys. {\bf A321} (1985) 467.\\
M.W. Kirson and A. Leviatan, Phys. Rev. Lett. {\bf 55} (1985) 2846.

\bibitem{GK}
J.N.~Ginocchio and M.W. Kirson, Nucl. Phys. {\bf A350} (1980) 31.

\bibitem{CA}
H.T.~Chen and A.~Arima, Phys. Rev. Lett. {\bf 51} (1983) 447.

\bibitem{ELL}
J.P. Elliott, Proc. Roy. Soc. {\bf A245} (1958) 128, 562.

\bibitem{VER}
J.D. Vergados,
Nucl. Phys. {\bf A111} (1968) 681.

\bibitem{SU3}
A. Arima and F. Iachello, Ann. Phys. {\bf 111} (1978) 201.

\bibitem{ISA}
P. Van Isacker, Phys. Rev. {\bf C27} (1983) 2447.
 
\bibitem{WAL}
N. Whelan and Y. Alhassid and A. Leviatan,
Phys. Rev. Lett. {\bf 71} (1993) 2208.

\end{thebibliography}
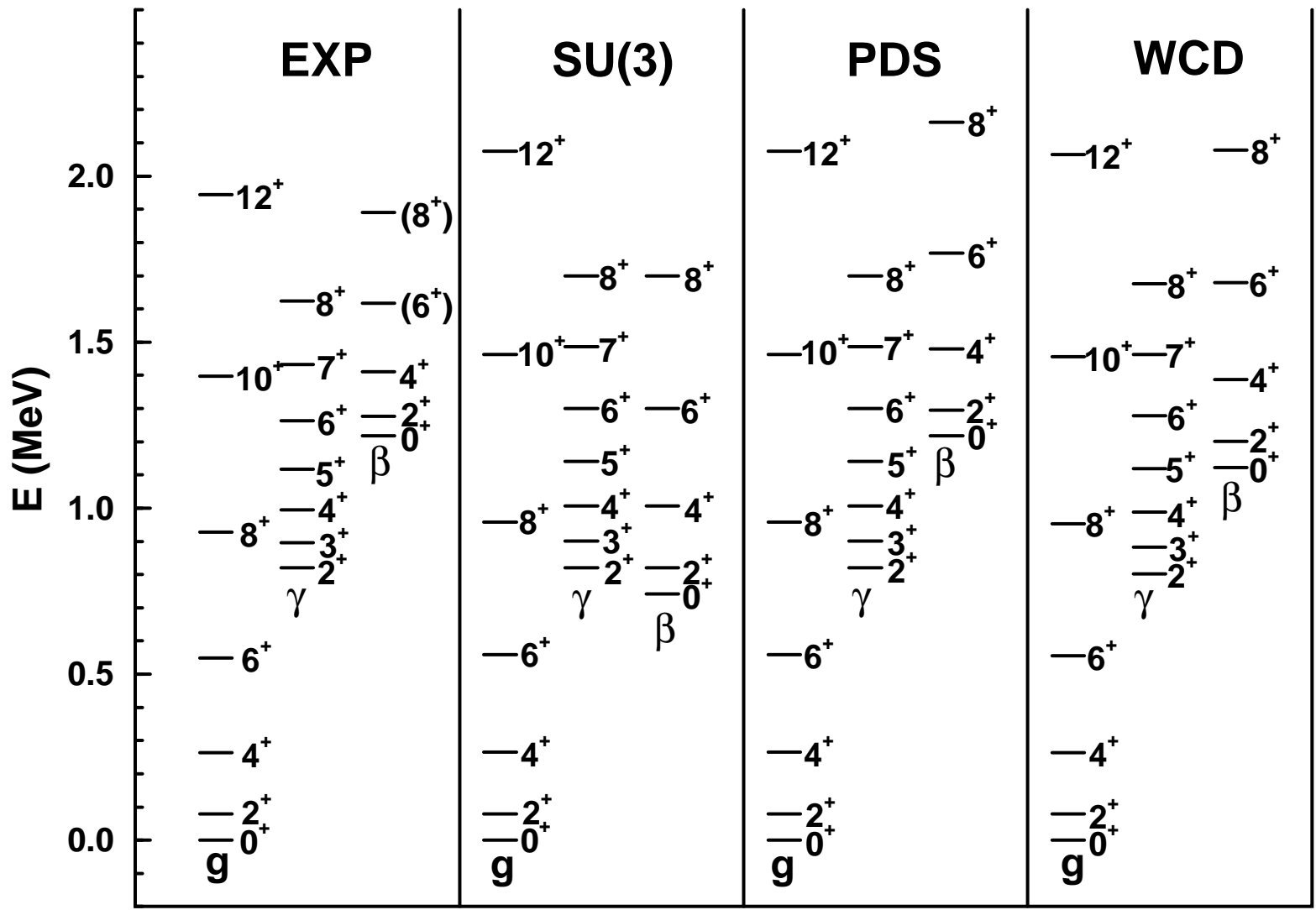
\end{document}